\documentclass[12pt,onecolumn]{revtex4}
\usepackage{amssymb}

\begin{document}

\title{Magnetic Strings in Dilaton Gravity}
\author{M. H. Dehghani}\email{mhd@shirazu.ac.ir}
\address{$1$. Physics Department and Biruni Observatory, Shiraz University, Shiraz 71454, Iran\\
         $2$. Research Institute for Astrophysics and Astronomy of Maragha (RIAAM), Maragha, Iran}

\begin{abstract}
First, I present two new classes of magnetic rotating solutions in
four-dimensional Einstein-Maxwell-dilaton gravity with Liouville-type
potential. The first class of solutions yields a $4$-dimensional spacetime
with a longitudinal magnetic field generated by a static or spinning
magnetic string. I find that these solutions have no curvature singularity
and no horizons, but have a conic geometry. In these spacetimes, when the
rotation parameter does not vanish, there exists an electric field, and
therefore the spinning string has a net electric charge which is
proportional to the rotation parameter. The second class of solutions yields
a spacetime with an angular magnetic field. These solutions have no
curvature singularity, no horizon, and no conical singularity. The net
electric charge of the strings in these spacetimes is proportional to their
velocities. Second, I obtain the ($n+1$)-dimensional rotating solutions in
Einstein-dilaton gravity with Liouville-type potential. I argue that these
solutions can present horizonless spacetimes with conic singularity, if one
chooses the parameters of the solutions suitable. I also use the counterterm
method and compute the conserved quantities of these spacetimes.
\end{abstract}


\maketitle
\section{Introduction\label{Intro}}

It seems likely, at least at sufficiently high energy scales, that gravity
is not governed by the Einstein's action, and is modified by the superstring
terms which are scalar tensor in nature. In the low energy limit of the
string theory, one recovers Einstein gravity along with a scalar dilaton
field which is non minimally coupled to the gravity \cite{Green}.
Scalar-tensor theories are not new, and it was pioneered by Brans and Dicke
\cite{Bran}, who sought to incorporate Mach's principle into gravity.

In this paper I want to obtain new exact horizonless solutions of
four-dimensional Einstein-Maxwell-dilaton gravity and higher
dimensional Einstein-dilaton gravity. There are many papers which
are dealing directly with the issue of spacetimes generated by
string source that are horizonless and have non trivial external
solutions. Static uncharged cylindrically symmetric solutions of
Einstein gravity in four dimensions with vanishing cosmological
constant have been considered in \cite{Levi}. Similar static
solutions in the context of cosmic string theory have been found
in \cite {Vil}. All of these solutions \cite{Levi,Vil} are
horizonless and have a conical geometry; they are everywhere flat
except at the location of the line source. An extension to include
the electromagnetic field has also been done \cite{Muk}.
Asymptotically anti de Sitter (AdS) spacetimes generated by static
and spinning magnetic sources in three and four dimensional
Einstein-Maxwell gravity with negative cosmological constant have
been investigated in \cite{Lem1}. The generalization of these
asymptotically AdS magnetic rotating solutions of the
Einstein-Maxwell equation to higher dimensions \cite{Deh1} and
higher derivative gravity \cite{Deh2} have also been done. In the
context of electromagnetic cosmic string, it was shown that there
are cosmic strings, known as superconducting cosmic string, that
behave as superconductors and have interesting interactions with
astrophysical magnetic fields \cite{Wit}. The properties of these
superconducting cosmic strings have been investigated in
\cite{Moss}. Superconducting cosmic strings have also been studied
in Brans-Dicke theory \cite{Sen1}, and in dilaton gravity
\cite{Fer}.

On the other side, some efforts have been done to construct exact
solutions of Einstein-Maxwell-dilaton gravity. Exact charged
dilaton black hole solutions in the absence of dilaton potential
have been constructed by many authors \cite{CDB1, CDB2}. In the
presence of Liouville potential, static charged black hole
solutions have also been discovered with a positive constant
curvature event horizons \cite{Hor2}, and zero or negative
constant curvature horizons \cite{Cai}. These exact solutions are
all static. Till now, charged rotating dilaton solutions for an
arbitrary coupling constant has not been constructed in four or
higher dimensions. Indeed, exact magnetic rotating solutions have
been considered in three dimensions \cite{Dia}, while exact
rotating black hole solutions in four dimensions have been
obtained only for some limited values of the coupling constant
\cite{Fr}. For general dilaton coupling, the properties of
rotating charged dilaton black holes only with infinitesimally
small angular momentum \cite{Hor1} or small charge \cite{Cas} have
been investigated, while for arbitrary values of angular momentum
and charge only a numerical investigation has been done
\cite{Klei}. My aim, here, is to construct exact rotating charged
dilaton solutions for an arbitrary value of coupling constant.

The outline of the paper is as follows. Section \ref{Cons} is devoted to a
brief review of the field equations and general formalism of calculating the
conserved quantities. In Sec \ref{Charged}, I obtain the four-dimensional
rotating dilaton strings which produce longitudinal and angular magnetic
fields, and investigate their properties. In Sec. \ref{Unch}, I present the $%
(n+1)$-dimensional rotating dilaton solutions with conic singularity and
compute their conserved charges. I finish the paper with some concluding
remarks.

\section{Field Equations and Conserved Quantities\label{Cons}}

The action of dilaton Einstein-Maxwell gravity with one scalar field $\Phi $
with Liouville-type potential in $(n+1)$ dimensions is \cite{Hor2}
\begin{eqnarray}
I_{G} &=&-\frac{1}{16\pi }\int_{\mathcal{M}}d^{n+1}x\sqrt{-g}\left( \mathcal{%
R}\text{ }-\frac{4}{n-1}(\nabla \Phi )^{2}-2\Lambda e^{2\beta \Phi
}-e^{-4\alpha \Phi/(n-1)}F_{\mu \nu }F^{\mu \nu }\right)  \nonumber \\
&&+\frac{1}{8\pi }\int_{\partial \mathcal{M}}d^{n}x\sqrt{-\gamma }\Theta
(\gamma ),  \label{Act}
\end{eqnarray}
where $\mathcal{R}$ is the Ricci scalar, $\alpha $ is a constant
determining the strength of coupling of the scalar and
electromagnetic fields, $F_{\mu \nu }=\partial _{\mu }A_{\nu
}-\partial _{\nu }A_{\mu }$ is the electromagnetic tensor field
and $A_{\mu }$ is the vector potential. The last term in Eq.
(\ref{Act}) is the Gibbons-Hawking boundary term. The
manifold $\mathcal{M}$ has metric $g_{\mu \nu }$ and covariant derivative $%
\nabla _{\mu }$. $\Theta $ is the trace of the extrinsic curvature $\Theta
^{\mu \nu }$ of any boundary(ies) $\partial \mathcal{M}$ of the manifold $%
\mathcal{M}$, with induced metric(s) $\gamma _{ij}$. One may refer to $%
\Lambda $ as the cosmological constant, since in the absence of the dilaton
field ($\Phi =0$) the action (\ref{Act}) reduces to the action of
Einstein-Maxwell gravity with cosmological constant.

The field equations in $(n+1)$ dimensions are obtained by varying the action
(\ref{Act}) with respect to the dynamical variables $A_{\mu }$, $g_{\mu \nu
} $ and $\Phi $:
\begin{eqnarray}
&&\partial _{\mu }\left[ \sqrt{-g}e^{-4\alpha \Phi/(n-1)}F^{\mu \nu }%
\right] =0,  \label{Fil1} \\
&&\mathcal{R}_{\mu \nu }=\frac{2}{n-1}\left( 2\partial _{\mu }\Phi
\partial_{\nu }\Phi +\Lambda e^{2\beta \Phi }g_{\mu \nu
}\right) +2e^{-4\alpha \Phi/(n-1)}\left( F_{\mu \lambda }F_{\nu }^{\text{ }\lambda }-\frac{%
1}{2(n-1)}F_{\rho \sigma }F^{\rho \sigma }g_{\mu \nu }\right),
\label{Fil2}
\\
&&\nabla ^{2}\Phi =\frac{n-1}{2}\beta \Lambda e^{2\beta \Phi
}-\frac{\alpha }{2}e^{-4\alpha \Phi/(n-1)}F_{\rho \sigma }F^{\rho
\sigma }. \label{Fil3}
\end{eqnarray}

The conserved mass and angular momentum of the solutions of the above field
equations can be calculated through the use of the substraction method of
Brown and York \cite{BY}. Such a procedure causes the resulting physical
quantities to depend on the choice of reference background. For
asymptotically (A)dS solutions, the way that one deals with these
divergences is through the use of counterterm method inspired by (A)dS/CFT
correspondence \cite{Mal}. However, in the presence of a non-trivial dilaton
field, the spacetime may not behave as either dS ($\Lambda >0$) or AdS ($%
\Lambda <0$). In fact, it has been shown that with the exception of a pure
cosmological constant potential, where $\beta =0$, no AdS or dS static
spherically symmetric solution exist for Liouville-type potential \cite{Pol}%
. But, as in the case of asymptotically AdS spacetimes, according to the
domain-wall/QFT (quantum field theory) correspondence \cite{Sken}, there may
be a suitable counterterm for the stress energy tensor which removes the
divergences. In this paper, I deal with the spacetimes with zero curvature
boundary [$R_{abcd}(\gamma )=0$], and therefore the counterterm for the
stress energy tensor should be proportional to $\gamma ^{ab}$. Thus, the
finite stress-energy tensor in $(n+1)$ dimensions may be written as
\begin{equation}
T^{ab}=\frac{1}{8\pi }\left[ \Theta ^{ab}-\Theta \gamma ^{ab}+\frac{n-1}{l_{%
\mathrm{eff}}}\gamma ^{ab}\right] ,  \label{Stres}
\end{equation}
where $l_{\mathrm{eff}}$ is given by
\begin{equation}
l_{\mathrm{eff}}^{2}=\frac{(n-1)^{3}\beta ^{2}-4n(n-1)}{8\Lambda }e^{-2\beta
\Phi }.  \label{leff}
\end{equation}
As $\beta $ goes to zero, the effective $l^2_{\mathrm{eff}}$ of Eq. (\ref{leff}%
) reduces to $l^2=-n(n-1)/2\Lambda $ of the AdS spacetimes. The
first two terms in Eq. (\ref{Stres}) is the variation of the
action (\ref{Act}) with respect to $\gamma ^{ab}$, and the last
term is the counterterm which removes the divergences. One may
note that the counterterm has the same form as in the case of
asymptotically AdS solutions with zero curvature boundary, where
$l$ is replaced by $l_{\mathrm{eff}}$. To compute the conserved
charges of the spacetime, one should choose a spacelike surface
$\mathcal{B}$ in $\partial \mathcal{M}$ with metric $\sigma
_{ij}$, and write the boundary metric in ADM form:
\[
\gamma _{ab}dx^{a}dx^{a}=-N^{2}dt^{2}+\sigma _{ij}\left( d\varphi
^{i}+V^{i}dt\right) \left( d\varphi ^{j}+V^{j}dt\right) ,
\]
where the coordinates $\varphi ^{i}$ are the angular variables
parameterizing the hypersurface of constant $r$ around the origin, and $N$
and $V^{i}$ are the lapse and shift functions respectively. When there is a
Killing vector field $\mathcal{\xi }$ on the boundary, then the quasilocal
conserved quantities associated with the stress tensors of Eq. (\ref{Stres})
can be written as
\begin{equation}
\mathcal{Q}(\mathcal{\xi )}=\int_{\mathcal{B}}d^{n-1}\varphi \sqrt{\sigma }%
T_{ab}n^{a}\mathcal{\xi }^{b},  \label{charge}
\end{equation}
where $\sigma $ is the determinant of the metric $\sigma _{ij}$, and $\mathcal{%
\xi }$ and $n^{a}$ are the Killing vector field and the unit normal vector
on the boundary $\mathcal{B}$ . For boundaries with timelike ($\xi =\partial
/\partial t$), rotational ($\varsigma =\partial /\partial \varphi $) and
translational Killing vector fields ($\zeta =\partial /\partial x$), one
obtains the quasilocal mass, angular and linear momenta
\begin{eqnarray}
M &=&\int_{\mathcal{B}}d^{n-1}\varphi \sqrt{\sigma }T_{ab}n^{a}\xi ^{b},
\label{Mastot} \\
J &=&\int_{\mathcal{B}}d^{n-1}\varphi \sqrt{\sigma }T_{ab}n^{a}\varsigma
^{b},  \label{Angtot} \\
P &=&\int_{\mathcal{B}}d^{n-1}\varphi \sqrt{\sigma }T_{ab}n^{a}\zeta ^{b},
\label{Lintot}
\end{eqnarray}
provided the surface $\mathcal{B}$ contains the orbits of $\varsigma $.
These quantities are, respectively, the conserved mass, angular and linear
momenta of the system enclosed by the boundary $\mathcal{B}$. Note that they
will both be dependent on the location of the boundary $\mathcal{B}$ in the
spacetime, although each is independent of the particular choice of
foliation $\mathcal{B}$ within the surface $\partial \mathcal{M}$.

\section{Four-dimensional Horizonless Dilaton Solutions \label{Charged}}

In this section I want to obtain the $4$-dimensional horizonless solutions
of Eqs. (\ref{Fil1})-(\ref{Fil3}). First, a spacetime generated by a
magnetic source which produces a longitudinal magnetic field along the $z$%
-axis is constructed, and second, I obtain a spacetime generated by a
magnetic source that produces angular magnetic fields along the $\varphi $
coordinate.

\subsection{The Longitudinal Magnetic Field Solutions\label{Long}}

I assume that the metric has the following form:
\begin{equation}
ds^{2}=-\frac{\rho ^{2}}{l^{2}}R^{2}(\rho )dt^{2}+\frac{d\rho ^{2}}{f(\rho )}%
+l^{2}f(\rho )d\varphi ^{2}+\frac{\rho ^{2}}{l^{2}}R^{2}(\rho )dz^{2},
\label{Met1a}
\end{equation}
where the constant $l$ have dimension of length which is related to the
cosmological constant $\Lambda $ in the absence of a dilaton field ($\Phi =0$%
). Note that the coordinate $z$ ( $-\infty <z<\infty )$\ has the dimension
of length, while the angular coordinate $\varphi $ is dimensionless as usual
and ranges in $0\leq \varphi <2\pi $. The motivation for this metric gauge $%
[g_{tt}\varpropto -\rho ^{2}$ and $(g_{\rho \rho })^{-1}\varpropto
g_{\varphi \varphi }]$ instead of the usual Schwarzschild gauge $[(g_{\rho
\rho })^{-1}\varpropto g_{tt}$ and $g_{\varphi \varphi }\varpropto \rho
^{2}] $ comes from the fact that I am looking for a magnetic solution
instead of an electric one.

The Maxwell equation (\ref{Fil1}) for the metric (\ref{Met1a}) is $\partial
_{\mu }\left[ \rho^{2}R^{2}(\rho )\exp (-2\alpha \Phi )F^{\mu \nu }\right]
=0,$ which shows that if one choose
\begin{equation}
R(\rho )=\exp (\alpha \Phi ),  \label{Rr}
\end{equation}
then the vector potential is

\begin{equation}
A_{\mu }=-\frac{ql}{\rho }\delta _{\mu }^{\varphi },  \label{met1b}
\end{equation}
where $q$ is the charge parameter. The field equations (\ref{Fil2}) and (\ref
{Fil3}) for the metric (\ref{Met1a}) can be written as:
\begin{eqnarray}
&&f^{^{\prime }}(1+\alpha \rho\Phi ^{^{\prime }})+\alpha f[\rho \Phi
^{^{\prime \prime }}+2\rho \alpha \Phi ^{\prime 2}+4\Phi ^{\prime }+(\alpha
\rho )^{-1}]-q^{2}\rho ^{-3}e^{-2\alpha \Phi }+\Lambda \rho e^{2\beta \Phi
}=0,  \label{E1} \\
&&\frac{\rho}{2} f^{^{\prime \prime }}+f^{^{\prime }}(1+\alpha \rho \Phi
^{^{\prime }})+2f[\alpha \rho \Phi ^{^{\prime \prime }}+\rho (1+\alpha
^{2})\Phi ^{\prime 2}+2\alpha \Phi ^{\prime }]+\frac{q^{2}}{\rho ^{3}}%
e^{-2\alpha \Phi }+\Lambda \rho e^{2\beta \Phi }=0,  \label{E2} \\
&&\frac{\rho}{2} f^{^{\prime \prime }}+f^{^{\prime }}(1+\alpha \rho \Phi
^{^{\prime }})+q^{2}\rho ^{-3}e^{-2\alpha \Phi }+\Lambda \rho e^{2\beta \Phi
}=0,  \label{E3} \\
&&f\Phi ^{^{\prime \prime }}+f^{^{\prime }}\Phi ^{^{\prime }}+2\alpha f\Phi
^{\prime 2}+2\rho ^{-1}f\Phi ^{\prime }+\alpha q^{2}\rho ^{-4}e^{-2\alpha
\Phi }-\beta \Lambda e^{2\beta \Phi }=0,  \label{E4}
\end{eqnarray}
where ``prime'' denotes differentiation with respect to\ $\rho $.
Subtracting Eq. (\ref{E2}) from Eq. (\ref{E3}) gives:
\[
\alpha \rho \Phi ^{\prime \prime }+2\alpha \Phi ^{\prime }+\rho (1+\alpha
^{2})\Phi ^{\prime 2}=0,
\]
which shows that $\Phi (\rho )$ can be written as:
\begin{equation}
\Phi (\rho )=\frac{\alpha }{1+\alpha ^{2}}\ln \left(\frac{b}{\rho }+c
\right),  \label{Ph}
\end{equation}
where $b$ and $c$ are two arbitrary constants. Using the expression (\ref{Ph}%
) for $\Phi (\rho )$ in Eqs. (\ref{E1})-(\ref{E4}), one finds that these
equations are inconsistent for $c\neq 0$. Thus, $c$ should vanish.

The only case that I find exact solutions for an arbitrary values of $%
\Lambda $ (including $\Lambda =0$) with $R(\rho )$ and $\Phi (\rho )$ of
Eqs. (\ref{Rr}) and (\ref{Ph}) is when $\beta =\alpha $. It is easy, then,
to obtain the function $f(\rho )$ as
\begin{equation}
f(\rho )=\rho ^{2\gamma }\left( \frac{\Lambda V_{0}(1+\alpha ^{2})^{2}}{%
\alpha ^{2}-3}\rho ^{2(1-2\gamma )}+\frac{m}{\rho }-\frac{(1+\alpha
^{2})q^{2}}{V_{0}\rho ^{2}}\right) ,  \label{Fr1a}
\end{equation}
where $\gamma =\alpha ^{2}/(1+\alpha ^{2})$ and $V_{0}=b^{2\gamma }$. In the
absence of a non-trivial dilaton ($\alpha =0=\gamma $), the solution reduces
to the asymptotically AdS horizonless magnetic string for $\Lambda =-3/l^{2}$
\cite{Lem1}. As one can see from Eq. (\ref{Fr1a}), there is no solution for $%
\alpha =\sqrt{3}$ with a Liouville potential ($\Lambda \neq 0$).

In order to study the general structure of this solution, one may first look
for curvature singularities. It is easy to show that the Kretschmann scalar $%
R_{\mu \nu \lambda \kappa }R^{\mu \nu \lambda \kappa }$ diverges at $\rho =0$
and therefore one might think that there is a curvature singularity located
at $\rho =0$. However, as will be seen below, the spacetime will never
achieve $\rho =0$. Second, one looks for the existence of horizons, and
therefore one searches for possible black hole solutions. The horizons, if
any exist, are given by the zeros of the function $f(\rho )=g^{\rho \rho }$.
Let us denote the smallest positive root of $f(\rho )=0$ by $r_{+}$. The
function $f(\rho )$ is negative for $\rho <r_{+}$, and therefore one may
think that the hypersurface of constant time and $\rho =r_{+}$ is the
horizon. However, this analysis is not correct. Indeed, one may note that $%
g_{\rho \rho }$ and $g_{\phi \phi }$ are related by $f(\rho )=g_{\rho \rho
}^{-1}=l^{-2}g_{\phi \phi }$, and therefore when $g_{\rho \rho }$ becomes
negative (which occurs for $\rho <r_{+}$) so does $g_{\phi \phi }$. This
leads to an apparent change of signature of the metric from $+2$ to $-2$,
and therefore indicates that an incorrect extension is used. To get rid of
this incorrect extension, one may introduce the new radial coordinate $r$ as
\begin{equation}
r^{2}=\rho ^{2}-r_{+}^{2}\Rightarrow d\rho ^{2}=\frac{r^{2}}{r^{2}+r_{+}^{2}}%
dr^{2}.  \label{Tr1}
\end{equation}
With this new coordinate, the metric (\ref{Met1a}) is
\begin{eqnarray}
ds^{2} &=&-\frac{r^{2}+r_{+}^{2}}{l^{2}}e^{2\alpha \Phi
}dt^{2}+l^{2}f(r)d\phi ^{2}  \nonumber \\
&&+\frac{r^{2}}{(r^{2}+r_{+}^{2})f(r)}dr^{2}+\frac{r^{2}+r_{+}^{2}}{l^{2}}%
e^{2\alpha \Phi }dz^{2},  \label{Met1b}
\end{eqnarray}
where $f(r)$ is now given as
\begin{equation}
f(r)=(r^{2}+r_{+}^{2})^{\gamma }\left( \frac{\Lambda V_{0}(1+\alpha ^{2})^{2}%
}{(\alpha ^{2}-3)}(r^{2}+r_{+}^{2})^{(1-2\gamma )}+\frac{m}{%
(r^{2}+r_{+}^{2})^{1/2}}-\frac{(1+\alpha ^{2})q^{2}}{V_{0}(r^{2}+r_{+}^{2})}%
\right) .  \label{Fr1b}
\end{equation}
The gauge potential in the new coordinate is
\begin{equation}
A_{\mu }=\frac{ql}{(r^{2}+r_{+}^{2})^{1/2}}\delta _{\mu }^{\varphi }.
\end{equation}

Now it is a matter of calculation to show that the Kretschmann scalar does
not diverge in the range $0\leq r<\infty $. However, the spacetime has a
conic geometry and has a conical singularity at $r=0$, since:
\[
\lim_{r\rightarrow 0}\frac{1}{r}\sqrt{\frac{g_{\varphi \varphi }}{g_{rr}}}=%
\frac{1}{2}mlr_{+}^{2(\gamma -1)}+\frac{2(1+\alpha ^{2})}{(\alpha ^{2}-3)}%
\Lambda lV_{0}r_{+}^{1-2\gamma }\neq 1.
\]
That is, as the radius $r$ tends to zero, the limit of the ratio ``\textrm{%
circumference/radius}'' is not $2\pi $ and therefore the spacetime has a
conical singularity at $r=0$. In order to investigate the casual structure
of the spacetime, I consider it for different ranges of $\alpha $ separately.

For $\alpha >\sqrt{3}$, as $r$ goes to infinity the dominant term in Eq. (
\ref{Fr1b}) is the second term, and therefore the function $f(r)$ is
positive in the whole spacetime, despite the sign of the cosmological
constant $\Lambda $, and is zero at $r=0$. Thus, the solution given by Eqs. (%
\ref{Met1b}) and (\ref{Fr1b}) exhibits a spacetime with conic singularity at
$r=0$.

For $\alpha <\sqrt{3}$, the dominant term for large values of $r$\ is the
first term, and therefore the function $f(r)$ given in Eq. (\ref{Fr1b}) is
positive in the whole spacetime only for negative values of $\Lambda $. In
this case the solution presents a spacetime with conic singularity at $r=0$.
The solution is not acceptable for $\alpha <\sqrt{3}$ with positive values
of $\Lambda $, since the function $f(r)$ is negative for large values of $r$.

Of course, one may ask for the completeness of the spacetime with $r\geq 0$
\cite{Hor3}. It is easy to see that the spacetime described by Eq. (\ref
{Met1b}) is both null and timelike geodesically complete for $r\geq 0$. To
do this, one may show that every null or timelike geodesic starting from an
arbitrary point either can be extended to infinite values of the affine
parameter along the geodesic or will end on a singularity at $r=0$. Using
the geodesic equation, one obtains
\begin{eqnarray}
&& \dot{t} =\frac{l^{2}}{V_{0}(r^{2}+r_{+}^{2})^{1-\gamma }}E,\hspace{0.5cm}%
\dot{z}=\frac{l^{2}}{V_{0}(r^{2}+r_{+}^{2})^{1-\gamma }}P,\hspace{0.5cm}\dot{%
\phi}=\frac{1}{l^{2}f(r)}L,  \label{Geo1} \\
&& r^{2}\dot{r}^{2} =(r^{2}+r_{+}^{2})f(r)\left[ \frac{l^{2}(E^{2}-P^{2})}{%
V_{0}(r^{2}+r_{+}^{2})^{1-\gamma }}-\kappa \right] -\frac{r^{2}+r_{+}^{2}}{%
l^{2}}L^{2},  \label{Geo2}
\end{eqnarray}
where the overdot denotes the derivative with respect to an affine
parameter, and $\kappa $ is zero for null geodesics and $+1$ for timelike
geodesics. $E$, $L$, and $P$ are the conserved quantities associated with
the coordinates $t$, $\phi $, and $z$, respectively. Notice that $f(r)$ is
always positive for $r>0$ and zero for $r=0$.

First, I consider the null geodesics ($\kappa =0$). (i) If $E>P$ the
spiraling particles ($L>0$) coming from infinity have a turning point at $%
r_{tp}>0$, while the nonspiraling particles ($L=0$) have a turning point at $%
r_{tp}=0$. (ii) If $E=P$ and $L=0$, whatever the value of $r$, $\dot{r}$ and
$\dot{\phi}$ vanish and therefore the null particles moves on the $z$-axis.
(iii) For $E=P$ and $L\neq 0$, and also for $E<P$ and any values of $L$,
there is no possible null geodesic.

Second, I analyze the timelike geodesics ($\kappa =+1$). Timelike
geodesics are possible only if
$l^{2}(E^{2}-P^{2})>V_{0}r_{+}^{2(1-\gamma )}$. In this
case the turning points for the nonspiraling particles ($L=0$) are $%
r_{tp}^{1}=0$ and $r_{tp}^{2}$ given as
\begin{equation}
r_{tp}^{2}=\sqrt{[V_{0}^{-1}l^{2}(E^{2}-P^{2})]^{1/(1-\gamma )}-r_{+}^{2}},
\end{equation}
while the spiraling ($L\neq 0$) timelike particles are bound between $%
r_{tp}^{a}$ and $r_{tp}^{b}$ given by $0<r_{tp}^{a}\leq
r_{tp}^{b}<r_{tp}^{2} $. Thus, I confirmed that the spacetime described by
Eq. (\ref{Met1b}) is both null and timelike geodesically complete.

\subsection{The Rotating Longitudinal Magnetic Field Solutions}

\label{Angul} Now, I want to endow the spacetime solution (\ref{Met1b}) with
a global rotation. In order to add angular momentum to the spacetime, one
may perform the following rotation boost in the $t-\phi $ plane
\begin{equation}
t\mapsto \Xi t-a\varphi ,\hspace{0.5cm}\varphi \mapsto \Xi \varphi -\frac{a}{%
l^{2}}t,  \label{Tr}
\end{equation}
where $a$ is a rotation parameter and $\Xi =\sqrt{1+a^{2}/l^{2}}$.
Substituting Eq. (\ref{Tr}) into Eq. (\ref{Met1b}) one obtains
\begin{eqnarray}
ds^{2} &=&-\frac{r^{2}+r_{+}^{2}}{l^{2}}e^{2\alpha \Phi }\left( \Xi
dt-ad\varphi \right) ^{2}+\frac{r^{2}dr^{2}}{(r^{2}+r_{+}^{2})f(r)}
\nonumber \\
&&+l^{2}f(r)\left( \frac{a}{l^{2}}dt-\Xi d\varphi \right) ^{2}+\frac{%
r^{2}+r_{+}^{2}}{l^{2}}e^{2\alpha \Phi }dz^{2},  \label{Met1c}
\end{eqnarray}
where $f(r)$ is the same as $f(r)$ given in Eq. (\ref{Fr1b}). The gauge
potential is now given by
\begin{equation}
A_{\mu }=-\frac{q}{(r^{2}+r_{+}^{2})^{1/2}}\left( \frac{a}{l}\delta _{\mu
}^{t}-\Xi l\delta _{\mu }^{\varphi }\right) .  \label{Pot2}
\end{equation}
The transformation (\ref{Tr}) generates a new metric, because it is not a
permitted global coordinate transformation \cite{Sta}. This transformation
can be done locally but not globally. Therefore, the metrics (\ref{Met1b})
and (\ref{Met1c}) can be locally mapped into each other but not globally,
and so they are distinct. Note that this spacetime has no horizon and
curvature singularity. However, it has a conical singularity at $r=0$. It is
notable to mention that this solution reduces to the solution of
Einstein-Maxwell equation introduced in \cite{Lem1} as $\alpha $ goes to
zero.

The mass and angular momentum per unit length of the string when the
boundary $\mathcal{B}$ goes to infinity can be calculated through the use of
Eqs. (\ref{Mastot}) and (\ref{Angtot}),
\begin{equation}
\mathcal{M}=\frac{V_{0}[(3-\alpha ^{2})\Xi ^{2}-2]}{8l(1+\alpha ^{2})}m,%
\hspace{0.5cm}\mathcal{J}=\frac{(3-\alpha ^{2})V_{0}}{8l(1+\alpha ^{2})}\Xi
ma.  \label{MJ}
\end{equation}
For $a=0$ ($\Xi =1$), the angular momentum per unit length vanishes, and
therefore $a$ is the rotational parameter of the spacetime. Of course, one
may note that these conserved charges reduce to the conserved charges of the
rotating black string obtained in Ref. \cite{Deh2} as $\alpha \rightarrow 0$.

\subsection{The Angular Magnetic Field Solutions\label{AngMag}}

In subsection \ref{Long}, I found a spacetime generated by a magnetic source
which produces a longitudinal magnetic field along the $z$-axis. Now, I want
to obtain a spacetime generated by a magnetic source that produces angular
magnetic fields along the $\varphi $ coordinate. Following the steps of
subsection \ref{Long} but now with the roles of $\varphi $ and $z$
interchanged, one can directly write the metric and vector potential
satisfying the field equations (\ref{E1})-(\ref{E4}) as
\begin{eqnarray}
ds^{2} &=&-\frac{r^{2}+r_{+}^{2}}{l^{2}}e^{2\alpha \Phi }dt^{2}+\frac{%
r^{2}dr^{2}}{(r^{2}+r_{+}^{2})f(r)}  \nonumber \\
&&+(r^{2}+r_{+}^{2})e^{2\alpha \Phi }d\varphi ^{2}+f(r)dz^{2},  \label{Met2a}
\end{eqnarray}
where $f(r)$ is given in Eq. (\ref{Fr1b}). The angular coordinate $\varphi $
ranges in $0\leq \varphi <2\pi $. The gauge potential is now given by
\begin{equation}
A_{\mu }=\frac{q}{(r^{2}+r_{+}^{2})^{1/2}}\delta _{\mu }^{z}.  \label{Pot3}
\end{equation}
The Kretschmann scalar does not diverge for any $r$ and therefore there is
no curvature singularity. The spacetime (\ref{Met2a}) is also free of conic
singularity.

To add linear momentum to the spacetime, one may perform the boost
transformation $[t\mapsto \Xi t-(v/l)z$, $z\mapsto \Xi x-(v/l)t$] in the $%
t-z $ plane and obtain
\begin{eqnarray}
ds^{2} &=&-\frac{r^{2}+r_{+}^{2}}{l^{2}}e^{2\alpha \Phi }\left( \Xi dt-\frac{%
v}{l}dz\right) ^{2}+f(r)\left( \frac{v}{l}dt-\Xi dz\right) ^{2}  \nonumber \\
&&+\frac{r^{2}dr^{2}}{(r^{2}+r_{+}^{2})f(r)}+(r^{2}+r_{+}^{2})e^{2\alpha
\Phi }d\varphi ^{2},  \label{Met2b}
\end{eqnarray}
where $v\ $is a boost parameter and $\Xi =\sqrt{1+v^{2}/l^{2}}$. The gauge
potential is given by

\begin{equation}
A_{\mu }=-\frac{q}{(r^{2}+r_{+}^{2})^{1/2}}\left( \frac{v}{l}\delta _{\mu
}^{t}-\Xi \delta _{\mu }^{z}\right) .  \label{Pot4}
\end{equation}
Contrary to transformation (\ref{Tr}), this boost transformation is
permitted globally since $z$ is not an angular coordinate. Thus the boosted
solution (\ref{Met2b}) is not a new solution. However, it generates an
electric field.

The conserved quantities of the spacetime (\ref{Met2b}) are the mass and
linear momentum along the $z$-axis given as

\[
\mathcal{M}=\frac{V_{0}[(3-\alpha ^{2})\Xi ^{2}-2]}{8l(1+\alpha ^{2})}m,%
\hspace{0.5cm}\mathcal{P}=\frac{(3-\alpha ^{2})V_{0}}{8l(1+\alpha ^{2})}\Xi
mv.
\]

Now, I calculate the electric charge of the solutions (\ref{Met1c}) and (\ref
{Met2b}) obtained in this section. To determine the electric field one
should consider the projections of the electromagnetic field tensor on
special hypersurfaces. The normal to such hypersurfaces for the spacetimes
with a longitudinal magnetic field is
\[
u^{0}=\frac{1}{N},\text{ \ }u^{r}=0,\text{ \ }u^{i}=-\frac{V^{i}}{N},
\]
and the electric field is $E^{\mu }=g^{\mu \rho }\exp (-2\alpha \Phi
)F_{\rho \nu }u^{\nu }$. Then the electric charge per unit length $\mathcal{Q%
}$ can be found by calculating the flux of the electric field at infinity,
yielding
\begin{equation}
\mathcal{Q}=\frac{(\Xi ^{2}-1)q}{2l}.  \label{chden}
\end{equation}
Note that the electric charge of the string vanishes, when the string has no
angular or linear momenta.

\section{The Rotating Solutions in Various Dimensions \label{Unch}}

In this section I look for the uncharged rotating solutions of
field equations (\ref{Fil1})-(\ref{Fil3}) in $n+1$ dimensions. I
first obtain the uncharged static solution and then generalize it
to the case of rotating solution with all the rotation parameters.

\subsection{Static Solutions}

I assume the metric has the following form
\begin{equation}
ds^{2}=-\frac{\rho ^{2}}{l^{2}}e^{2\beta \Phi }dt^{2}+\frac{d\rho ^{2}}{%
f(\rho )}+l^{2}f(\rho )d\phi ^{2}+\frac{\rho ^{2}}{l^{2}}e^{2\beta \Phi
}d\Omega ^{2},  \label{Met3a}
\end{equation}
where $d\Omega ^{2}$ is the metric of the $(n-1)$-dimensional
hypersurface which has zero curvature. The field equations
(\ref{Fil1})-(\ref{Fil3}) in the absence of electromagnetic field
become
\begin{eqnarray}
&&f^{^{\prime }}(1+\beta \rho \Phi ^{^{\prime }})+\beta f[\rho \Phi
^{^{\prime \prime }}+(n-1)(2\Phi ^{^{\prime }}+\beta \rho \Phi ^{^{\prime
}2})+(n-2)(\beta \rho )^{-1}]+\frac{2\Lambda \rho }{n-1}e^{2\beta \Phi }=0,
\label{EE1} \\
&&\frac{\rho }{n-1}f^{^{\prime \prime }}+f^{^{\prime }}(1+\beta \rho \Phi
^{^{\prime }})+\frac{4\Lambda \rho }{(n-1)^{2}}e^{2\beta \Phi }  \nonumber \\
&&\hspace{1.6cm}+2(n-2)\rho f\left[ \beta \rho \Phi ^{^{\prime \prime
}}+2\beta \Phi ^{^{\prime }}+\rho \left( \beta ^{2}+\frac{4}{(n-1)^{2}}%
\right) \Phi ^{^{\prime 2}}\right] =0,  \label{EE2} \\
&&\frac{\rho }{n-1} f^{^{\prime \prime }}+f^{^{\prime }}(1+\beta \rho \Phi
^{^{\prime }})+\frac{4\Lambda \rho }{(n-1)^{2}} e^{2\beta \Phi }=0,
\label{EE3} \\
&&f\Phi ^{^{\prime \prime }}+f^{^{\prime }}\Phi ^{^{\prime }}+(n-1)\left[
\beta f\Phi ^{^{\prime 2}}+\rho ^{-1}f\Phi ^{^{\prime }}-\frac{1}{2}\beta
\Lambda e^{2\beta \Phi }\right] =0.  \label{EE4}
\end{eqnarray}
Subtracting Eq. (\ref{EE2}) from Eq. (\ref{EE3}) gives
\[
\beta \rho \Phi ^{^{\prime \prime }}+2\beta \Phi ^{^{\prime }}+\rho \left(
\beta ^{2}+\frac{4}{(n-1)^{2}}\right) \Phi ^{^{\prime 2}}=0,
\]
which shows that $\Phi (\rho )$ can be written as:
\begin{equation}
\Phi (\rho )=\frac{(n-1)^2\beta }{4+(n-1)^{2}\beta ^{2}}\ln \left(\frac{c}{%
\rho }+d \right),  \label{Ph1}
\end{equation}
where $c$ and $d$ are two arbitrary constants. Substituting $\Phi (\rho )$
of Eq. (\ref{Ph1}) into the field equations (\ref{EE1})-(\ref{EE4}), one
finds that they are consistent only for $d=0$. Putting $d=0$, then $f(\rho )$
can be written as
\begin{equation}
f(\rho ) =\frac{8\Lambda V_{0}}{(n-1)^{3}\beta ^{2}-4n(n-1)}\rho ^{2\Gamma
}+m\rho ^{1-(n-1)\Gamma },  \label{Fr3a}
\end{equation}
where
\begin{equation}
V_{0}=\Gamma ^{-2}c^{2(1-\Gamma )} ,\hspace{.5cm}\Gamma = 4\{(n-1)^{2}\beta
^{2}+4\}^{-1}.  \label{Fr3b}
\end{equation}
One may note that there is no solution for $(n-1)^{2}\beta ^{2}-4n=0$. For
the two cases of $(n-1)^{2}\beta ^{2}-4n>0$ with positive $\Lambda $, and $%
(n-1)^{2}\beta ^{2}-4n<0$\ with negative $\Lambda $, the function $f(\rho )$
is positive in the whole spacetime. In these two cases, since the
Kretschmann scalar diverges at $\rho =0$ and $f(\rho )$ is positive for $%
\rho >0$, the spacetimes present naked singularities. For $(n-1)^{2}\beta
^{2}-4n<0$ with positive $\Lambda $, the function $f(\rho )$ is negative for
$\rho >r_{c}$, where $r_{c}$ is the root of $\ f(\rho )=0$. In this case,
the signature of the metric (\ref{Met3a}) will be changed as $\rho $ becomes
larger than $r_{c}$, and therefore this case is not acceptable.

The only case that there exist horizonless solutions with conic singularity
is when $(n-1)^{2}\beta ^{2}-4n>0$ with negative $\Lambda $. In this case,
as $r$ goes to infinity the dominant term is the second term of Eq. (\ref
{Fr3a}), and therefore $f(\rho )>0$ for $\rho >r_{+}$, where $r_{+}$ is the
root of $\ f(\rho )=0$. Again, one should perform the transformation (\ref
{Tr1}). Thus, the horizonless solution can be written as
\begin{eqnarray}
ds^{2} &=&-\frac{r^{2}+r_{+}^{2}}{l^{2}}e^{2\beta \Phi
}dt^{2}+l^{2}f(r)d\phi ^{2}  \nonumber \\
&&+\frac{r^{2}}{(r^{2}+r_{+}^{2})f(r)}dr^{2}+\frac{r^{2}+r_{+}^{2}}{l^{2}}%
e^{2\beta \Phi }dz^{2},  \label{Met3b} \\
f(r) &=&\frac{8\Lambda V_{0}}{(n-1)^{3}\beta ^{2}-4n(n-1)}%
(r^{2}+r_{+}^{2})^{\Gamma }+m(r^{2}+r_{+}^{2})^{\frac{1}{2}[1-(n-1)\Gamma ]},
\label{Fr3c}
\end{eqnarray}
where $\Gamma $ and $V_{0}$ are given in Eq. (\ref{Fr3b}). Again this
spacetime has no horizon and no curvature singularity. However, it has a
conical singularity at $r=0$. One should note that this solution reduces to
the $(n+1)$-dimensional uncharged solution of Einstein gravity given in \cite
{Deh1} for $\beta =0$ ($\Gamma =1=V_{0}$).

\subsection{Rotating solutions with all the rotation parameters}

The rotation group in $(n+1)$-dimensions is $SO(n)$ and therefore the number
of independent rotation parameters for a localized object is equal to the
number of Casimir operators, which is $[n/2]\equiv k$, where $[n/2]$ is the
integer part of $n/2$. The generalization of the metric (\ref{Met3b}) with
all rotation parameters is
\begin{eqnarray}
ds^{2} &=&-\frac{r^{2}+r_{+}^{2}}{l^{2}}e^{2\beta \Phi }\left( \Xi dt-{{%
\sum_{i=1}^{k}}}a_{i}d\phi ^{i}\right) ^{2}+f(r)\left( \sqrt{\Xi ^{2}-1}dt-%
\frac{\Xi }{\sqrt{\Xi ^{2}-1}}{{\sum_{i=1}^{k}}}a_{i}d\phi ^{i}\right) ^{2}
\nonumber \\
&&+\frac{r^{2}dr^{2}}{(r^{2}+r_{+}^{2})f(r)}+\frac{r^{2}+r_{+}^{2}}{%
l^{2}(\Xi ^{2}-1)}e^{2\beta \Phi }{\sum_{i<j}^{k}}(a_{i}d\phi
_{j}-a_{j}d\phi _{i})^{2}+\frac{r^{2}+r_{+}^{2}}{l^{2}}e^{2\beta \Phi
}dX^{2},  \nonumber \\
\Xi ^{2} &=& 1+{\sum_{i=1}^{k}}\frac{a_{i}^{2}}{l^{2}} ,
\end{eqnarray}
where $a_{i}$'s are $k$ rotation parameters, $f(r)$ is given in Eq. (\ref
{Fr3c}), and $dX^{2}$ is now the Euclidean metric on the $(n-k-1)$%
-dimensional submanifold with volume $V_{n-k-1}$.

The conserved mass and angular momentum per unit volume $V_{n-k-1}$ of the
solution calculated on the boundary $\mathcal{B}$ at infinity can be
calculated through the use of Eqs. (\ref{Mastot}) and (\ref{Angtot}),
\begin{eqnarray}
\mathcal{M} &=&(2\pi )^{k}\frac{{\Gamma }^{n}{V_{0}}^{(n-1)/2}}{16\pi
l^{n-k-1}}\left\{ \left[ n-\frac{(n-1)^{2}\beta ^{2}}{4}\right] \Xi
^{2}-(n-1)\right\} m, \\
\hspace{0.5cm}\mathcal{J}_{i} &=&(2\pi )^{k}\frac{{\Gamma }^{n}{V_{0}}%
^{(n-1)/2}}{16\pi l^{n-k-1}}\left\{ n-\frac{(n-1)\beta ^{2}\Xi ^{2}}{4}%
\right\} \Xi ma_{i}.
\end{eqnarray}
Note that for Einstein gravity the above computed conserved quantities
reduce to those given in \cite{Deh1}.

\section{Closing Remarks}

Till now, no explicit rotating charged dilaton solutions have been found
except for some dilaton coupling such as $\alpha =\sqrt{3}$ \cite{Fr} or $%
\alpha =1$ when the string three-form $H_{abc}$ is included
\cite{Sen}. For general dilaton coupling, the properties of
charged dilaton black holes have been investigated only for
rotating solutions with infinitesimally small angular momentum
\cite{Hor1} or small charge \cite{Cas}. It has also been shown
numerically that for the case of rotating solutions $\alpha
=\sqrt{3}$ is a critical value while for larger values of $\alpha$
new effect can appear \cite{Klei}. In this paper I obtained two
classes of exact horizonless rotating solutions with
Liouville-type potentials in four dimensions, provided $\beta
=\alpha \neq \sqrt{3}$. These solutions are neither asymptotically
flat nor (A)dS. The first class of solutions yields a
$4$-dimensional spacetime with a longitudinal magnetic field [the
only nonzero component of the vector potential is $A_{\varphi }(r)
$] generated by a static magnetic string. I also found the
rotating spacetime with a longitudinal magnetic field by a
rotational boost transformation. These solutions have no curvature
singularity and no horizons, but have conic singularity at $r=0$.
In these spacetimes, when the rotation parameter is zero (static
case), the electric field vanishes, and therefore the string has
no net electric charge. For the spinning string, when the rotation
parameter is nonzero, the string has a net electric charge density
which is proportional to the rotation parameter $a$. The second
class of solutions yields a spacetime with angular magnetic field.
These solutions have no curvature singularity, no horizon, and no
conic singularity. Again, I found that the string in these
spacetimes have no net electric charge when the boost parameter
vanishes. I also showed that, for the case of the traveling string
with nonzero boost parameter, the net electric charge density of
the string is proportional to the magnitude of its velocity ($v$).
These solutions reduce to the magnetic rotating string of
\cite{Lem1} as $\alpha \rightarrow 0$. I also computed the
conserved quantities of the four-dimensional magnetic string
through the use of the counterterm method.

Next, I generalized the rotating uncharged solutions to arbitrary $n+1$
dimensions with all rotation parameters. These spacetimes present naked
singularities for $[(n-1)^{2}\beta ^{2}-4n<0$, $\Lambda <0]$ and $%
[(n-1)^{2}\beta ^{2}-4n>0$, $\Lambda >0]$. The only case that the spacetime
exhibits a horizonless dilaton string with conic singularity is when $%
(n-1)^{2}\beta ^{2}-4n>0$ and $\Lambda <0$.

Note that the $(n+1)$-dimensional rotating solutions obtained here
are uncharged. Thus, it would be interesting if one can construct
rotating solutions in $(n+1)$ dimensions in the presence of
dilaton and electromagnetic fields. One may also attempt to
generalize these kinds of solutions obtained here to the case of
two-term Liouville potential \cite
{Hor2}. The case of charged rotating black string in four dimensions and $%
(n+1)$-dimensional rotating black branes will be present
elsewhere.
\acknowledgments{This work has been supported by
Research Institute for Astronomy and Astrophysics of Maragha,
Iran}

\end{document}